\documentclass[aps,pre,floatfix,twocolumn,showpacs,amsmath,amssymb,superscriptaddress,10pt]{revtex4-1}

\usepackage{amsthm}
\usepackage{bm}
\usepackage{epsfig}

\usepackage{verbatim}
\usepackage{graphicx}
\usepackage{times}
\usepackage{subfigure}

\newcommand{\beq}{\begin{equation}}
\newcommand{\eeq}{\end{equation}}
\newcommand{\beqa}{\begin{eqnarray}}
\newcommand{\eeqa}{\end{eqnarray}}

\newcommand{\RR}{{\mathbb R}}
\newcommand{\CC}{{\mathbb C}}

\usepackage{xcolor}

\begin{document}

\title{Energy Landscape of the Finite-Size Mean-field 2-Spin Spherical Model and Topology Trivialization}

\author{Dhagash Mehta}
\email{dmehta@nd.edu}
\affiliation{Department of Applied and Computational Mathematics and Statistics, University of Notre Dame, Notre Dame, IN 46565, USA}
\affiliation{Department of Chemistry, The University of Cambridge, Cambridge CB2 1EW, UK.}

\author{Jonathan D. Hauenstein}
\email{hauenstein@nd.edu}
\affiliation{Department of Applied and Computational Mathematics and Statistics, University of Notre Dame, Notre Dame, IN 46565, USA}
\affiliation{Simons Institute for the Theory of Computing, University of California, Berkeley, CA 94720-2190, USA}

\author{Matthew Niemerg}
\email{research@matthewniemerg.com}
\affiliation{Simons Institute for the Theory of Computing, University of California, Berkeley, CA 94720-2190, USA}
\affiliation{National Institute of Mathematical Sciences, Daejeon, Korea}

\author{Nicholas J. Simm}
\email{n.simm@qmul.ac.uk}
\affiliation{School of Mathematical Sciences, Queen Mary, University of London, Mile End Road, E1 4NS, UK.}

\author{Daniel A. Stariolo}
\email{daniel.stariolo@ufrgs.br}
\affiliation{Instituto de F\'isica, Universidade Federal do Rio Grande do Sul and
National Institute of Science and Technology for Complex Systems, CP 15051, 91501-970 Porto Alegre, RS, Brasil.}

\begin{abstract}
\noindent
Motivated by the recently observed phenomenon of topology trivialization of potential energy landscapes (PELs) for several statistical mechanics
models, we perform a numerical study of the finite size $2$-spin spherical model using 
both numerical polynomial homotopy continuation 
and a reformulation via non-hermitian matrices.
The continuation approach computes all of the complex stationary points of this model
while the matrix approach computes the real stationary points.
Using these methods, we compute the average number
of stationary points while changing the topology of the PEL
as well as the variance.  Histograms of these stationary points
are presented along with an analysis regarding the complex stationary points. 
This work connects topology trivialization to
two different branches of mathematics: algebraic geometry and catastrophe theory,
which is fertile ground for further interdisciplinary research. 
\end{abstract}

\maketitle

%-----------------------------------------------------
\section{Introduction}
\label{intro}
Recently, in two independent studies, it was observed that the mean number of real stationary points of 
a certain class of statistical models changes drastically when changing a certain parameter $\mu$ 
%of the model (changed it to "a certain parameter" NS 19/09)
\cite{Kastner:2011zz,Mehta:2012qr,Fyo04,
Fyodorov-topo-triv2013, Fyodorov-lectures}. 
%Results of this nature go back to the article \cite{Fyo04} which enumerated the critical points of a random Gaussian landscape depending on a 
%control parameter $\mu$. 
It was shown that as $\mu$ tends to a critical value $\mu_{c}$, one observes a sharp phase transition, separating a region of exponential 
proliferation of critical points from one of only finitely many.

Furthermore, in Refs.~\cite{Kastner:2011zz,Mehta:2012qr,Mehta:2014comm}, the coupling parameter
of the nearest-neighbour $\phi^4$-model on the $2$-dimensional lattice was continuously varied and found that the number of real stationary points
changed from around $10^{8}$ to $O(1)$ for the $4\times 4$ lattice case. Independently,
in Ref.~\cite{Fyodorov-topo-triv2013}, the problem of computing the real stationary 
points of the function $E_{h}(\textbf{x}) = -\frac{1}{2}\textbf{x}^{T}H \textbf{x} - \textbf{h}^{T} \textbf{x}$ was considered. Here,
$\textbf{x}=\{x_1,\dotsc,x_N\}$ are $N$ real variables subject to the spherical constraint $\sum_{i=1}^{N}x_{i}^2 = N$,
$H$ is a random matrix from the Gaussian Orthogonal Ensemble (GOE) and $\textbf{h}$ is a vector whose entries are i.i.d. random 
variables with zero mean and variance $\sigma^2$. It was shown that the mean number of real stationary points of $E_{h}(\textbf{x})$
can vary from $2N$ to 2. In between these two extreme cases, 
two non-trivial regimes were identified:  first, when $\sigma \sim O(N^{-1/2})$,
the number of stationary points is
of order $N$ and second, when  $\sigma \sim O(N^{-1/6})$, the number of solutions is of order one. This gradual decrease
of the complexity of the random manifold was termed {\em topology trivialization}. A similar phenomenon is also recently reported in random
dynamical systems \cite{wainrib2013topological}.

In Ref.~\cite{Fyodorov-lectures},
the results were extended to a generalized class of models, namely, to the $p$-spin spin glass model defined on the sphere and a model
of a Gaussian landscape in a confining parabolic potential.  Interestingly, in the $p$-spin model with $p>2$, which naturally generalizes
the $p=2$ case, there exists a critical value of $\sigma = \sigma_c$ such that for $\sigma < \sigma_c$ the landscape \cite{Wales:04,RevModPhys.80.167} has an exponentially large number
of stationary points.  For $\sigma > \sigma_c$, the landscape behaves in much the same way as in the $p=2$ case, i.e., it is
possible to find two different scaling regimes with system size interpolating between a region with a large number of stationary
points and a final region with only two. The abrupt change in the number of stationary points at $\sigma_c$ can be formally related
to a thermodynamic phase transition in the Statistical Mechanics version of the model. 
In the same work, the author also shows similar results for a random Gaussian landscape with a parabolic non-random confinement. 
Surprisingly, this model behaves in a qualitatively similar way as the p-spin model. Nevertheless, the parameter which triggers
the topology trivialization effect is not an external field but a parameter related to the curvatures of the confining potential
and the Gaussian manifold. %sp - appears -> appear 17/09
A unifying methodology of these works was to relate the properties of the mean number of stationary points and
also of extrema (minima and maxima) of Gaussian manifolds to known properties of the eigenvalue distributions of random matrices, 
specifically of matrices belonging to the GOE.

In this work, we use two different numerical algorithms to compute several quantities related to the topology trivialization
scenario in the 2-spin spin glass model with a spherical constraint. The Numerical Polynomial Homotopy Continuation Method~\cite{SW:05,BHSW13,Mehta:2011xs} 
allows us to compute all the complex stationary points of a polynomial function. This enables us to make an exhaustive search of the (complex) stationary points.  
We also use a method
based on a link between the 2-spin spherical model and non-Hermitian random matrices. This second method, which does not readily
generalize to $p>2$, only computes the real stationary points
and allows for larger $N$.
%I added "the mean in terms of the" below - basically to emphasize that what's interesting is that the *mean* can be computed exactly, not (so much) the density of eigenvalues. NS - 18/09/2014.
In particular, we 
present results for the mean number of real stationary points 
for finite system sizes.  Interestingly, there exists in the literature
analytic results for this quantity in terms of the density of eigenvalues of the GOE for any finite $N$~\cite{Forrester2012}. Our numerical
results are in agreement with the predictions of analytic calculations for finite $N$, and we also show how the results
approach the asymptotic prediction in the limit $N \to \infty$. 
In particular, our computations verify the existence
of the two scaling regimes predicted in \cite{Fyodorov-topo-triv2013}. We also present calculations for the variance of the number
of stationary points as a function of scaling parameters characterizing the two regimes of topology trivialization together with
results for the full probability distributions.  To the best of our knowledge, no theoretical results exist predicting the behavior of these
quantities. 

We also use our methods to obtain rather detailed statistics on the global minimum of $E_{h}(\mathbf{x})$. The distribution of this random variable was investigated heuristically in \cite{Fyodorov-topo-triv2013} using the powerful technique of replicas. The authors obtained a prediction for the large deviations function of the distribution of $E_{\min}$, valid for $N \gg 1$ and up to some critical value of the energy $E_{c}$. This later inspired the recent work of Dembo and Zeitouni \cite{DemZei14} who rigorously derived a different large deviations formula for $E_{\min}$. Although the latter formula largely confirms the heuristic predictions of \cite{Fyodorov-topo-triv2013}, it revealed a small interval of energies near $E_{c}$ where the corresponding rate functions are actually different. Remarkably, it turns out that the difference between the two rate functions is small enough to be virtually undetectable from a numerical point of view. Our numerical results show
good agreement with the large deviations predictions in the region where these are valid. 

In the last section we address the computation of all the complex solutions in the different regimes of interest. This clearly show how as
the topology of the landscape becomes simpler a corresponding growth of the imaginary parts of the solutions emerge. 

%-----------------------------------------------------
\section{The Mean-field $2$-spin Spherical Model}
The $2$-spin spherical model is defined by the Hamiltonian or energy function:
\beq\label{2hamilton}
E_{h}(\textbf{x}) = -\frac{1}{2}\textbf{x}^{T}H \textbf{x} - \textbf{h}^{T} \textbf{x},
\eeq
where $\textbf{x}=(x_1,\dotsc,x_N)\in\RR^N$ is a set of $N$ real degrees of freedom subject to the spherical constraint
\beq\label{constraint}
\sum_{i=1}^{N}x_{i}^2 = N
\eeq
which restricts $\textbf{x}$ to lie on an $(N-1)$-sphere of radius $\sqrt{N}$. 

The coupling constants $H$ are $N\times N$ real symmetric matrices with elements $H_{ij}$ independently drawn from a
Gaussian distribution with zero mean and variance $\langle H_{ij}^2\rangle=J^{2}/N$ for $i< j$ and diagonal elements 
 with zero mean and variance $\langle H_{ii}^2\rangle=2J^2/N$. The external field $\textbf{h}$ is a real
random vector with each entry independently drawn from a Gaussian distribution with zero mean and variance $\sigma^2$.

In order to derive the equations for the stationary points of the energy, it is convenient to introduce
a Lagrange multiplier~$\lambda$. With the spherical constraint and the energy function,
we obtain the Lagrangian function:
\begin{equation}
 E(\textbf{x},\lambda) = E_{h}(\textbf{x}) + \lambda \left(- N + \sum_{i=1}^{N} x_i^2\right).
\end{equation}
The stationary points of the energy are defined by the system of $N+1$ equations:
\begin{eqnarray}
 \frac{\partial E(\textbf{x},\lambda)}{\partial x_{i}} &=& -\sum_{j=1}^N H_{ij} x_{j} - h_i + 2 \lambda x_{i} = 0, \mbox{ $i = 1,\dots, N$,}\nonumber \\
 \frac{\partial E(\textbf{x},\lambda)}{\partial \lambda} &=& \sum_{i=1}^{N}x_{i}^{2} - N = 0. \label{eq:stationary_eqs}
\end{eqnarray}

\subsection{Known Results}
In \cite{Fyodorov-topo-triv2013}, the authors identified two scaling regimes as a function of the intensity of the
external field. The first regime is observed when $\sigma^2 \propto N^{-1}$. In this regime, for any finite 
$\gamma=N\frac{\sigma^2}{2J^2}$, the mean number of real solutions of the stationary equations is of the order of ${\cal N}(\gamma)
\sim O(N)$, i.e. the system has a large number of solutions, if $N$ is large. An explicit expression for ${\cal N}(\gamma)$ was
obtained in the asymptotic limit $N \to \infty$, equations $(12)$ and $(13)$  in \cite{Fyodorov-topo-triv2013}. The second
scaling regime is observed when $\sigma^2 \propto N^{-1/3}$. In this regime, it is useful to introduce another control parameter
$\kappa=N^{1/3}\frac{\sigma^2}{J^2}$. Then, for any fixed $\kappa$, the number of real solutions turns out to be
of order ${\cal N}(\kappa) \sim O(1)$. As $\kappa$ increases without bound, the number of stationary points converges to $2$.  This is the minimal possible number of 
real solutions, and these correspond to a unique maximum and a minimum.
One sees this phenomena occur in both the $\gamma$ and $\kappa$ regimes, i.e., the number of
solutions gradually diminishes until the energy function has a single minimum and a maximum. This process, driven by the strength
of an external field applied to the system, is called {\em topology trivialization} \cite{Fyodorov-topo-triv2013,Fyodorov-lectures}.
While analytical approaches are usually limited to large-N calculations, an exact expression for the real number of stationary points
of processes in the $GOE$ ensemble is known {\em for any N} \cite{Auffinger2013,Fyodorov-topo-triv2013}:
\beqa
{\mathcal N}&=&2N \left(\frac{2(J^2+\sigma^2)}{2J^2+\sigma^2}\right)^{1/2}\left(\frac{J^2}{J^2+\sigma^2}\right)^{N/2}\ \times \nonumber \\
  && \int_{-\infty}^{\infty}{\mathbb E}_{GOE}\{\rho_N(\lambda)\}e^{\frac{N \sigma^2}{2(2J^2+\sigma^2)}\lambda^2} d\lambda,
\label{eq:mean.real.sols.exact}
\eeqa
where ${\mathbb E}_{GOE}\{\rho_N(\lambda)\}$ is the mean eigenvalue density of the $GOE$ ensemble for which there are exact expressions
for arbitrary $N$ in terms of Hermite polynomials \cite{Forrester2012}. We compared our exact numerical results for finite $N$ with
this expression in each regime. %sp - "Is is" -> "It is" and "asympotitc" -> "asymptotic" 17/09
It is also of interest to compare numerical results for finite $N$ with the asymptotic result obtained
in~\cite{Fyodorov-topo-triv2013}. In the $N \to \infty$ limit, the mean eigenvalue density of the $GOE$ ensemble leads to the
well known semicircular law. Then, it is easy to obtain the resulting limit of expression (\ref{eq:mean.real.sols.exact}). In the 
$\gamma$ regime, it reduces to:
\beq
\lim_{N \to \infty}\frac{{\mathcal N}}{2N}={\mathcal N}(\gamma)=e^{-\gamma}\int_{-\sqrt{2}}^{\sqrt{2}} \sqrt{2-\lambda^2}\,e^{\frac{\gamma}{2}\lambda^2} \frac{d\lambda}{\pi}
\label{eq:mean.real.sols.asymp.gamma}
\eeq
which is equation (12) in~\cite{Fyodorov-topo-triv2013}. In the $\kappa$ regime, the integral in~(\ref{eq:mean.real.sols.exact}) is
dominated, in the large $N$ limit, by the edge of the mean eigenvalue density, $\rho_{edge}$. Performing the limit as $N \to \infty$ while keeping $\kappa$ finite,
one arrives at the asymptotic expression for the mean number of solutions in this regime:
\beq
\lim_{N \to \infty} {\mathcal N}(\kappa)=4\,e^{-\kappa^3/24}\,\int_{-\infty}^{\infty}e^{\frac{\kappa}{2}z}\rho_{edge}(z)\,dz
\label{eq:mean.real.sols.asymp.kappa}
\eeq
as given by equation (15) in~\cite{Fyodorov-topo-triv2013}. 

%We also define $\gamma:= N\sigma^2/(2 J^2)$ and $\kappa:= N^{\frac{1}{3}} \sigma^2/J^2$ for the later convenience.

\section{The Numerical Polynomial Homotopy Method Specialized for the $2$-spin Model}

One approach for computing all of the stationary points of the $2$-spin model is
by solving a system of multivariate polynomial equtions using
the numerical polynomial homotopy continuation (NPHC) method
\cite{Mehta:2009,Mehta:2009zv,Mehta:2011xs,Mehta:2011wj,Kastner:2011zz,Maniatis:2012ex,Mehta:2012wk,Hughes:2012hg,Mehta:2012qr,
MartinezPedrera:2012rs,He:2013yk,Mehta:2013fza,Mehta:2014ura,Greene:2013ida,SW:05,BHSW13}.
In particular, in Refs.~\cite{Greene:2013ida,Mehta:2013fza,Hughes:2012hg}, the method was used to explore the potential energy landscapes
of different potentials with random disorders, and in 
Ref.~\cite{hauenstein2014experiments} in a different statistical setting. 
The NPHC method can find \textit{all} the isolated complex solutions of the system
(see e.g.~\cite{KowalskiJ98,ThomH08,pielaks89} for related approaches).
It works by first determining an upper bound on the number of isolated
complex solutions of the given system. 
One such upper bound is the B\'ezout bound, which is simply 
the product of the degree of each polynomial equation. 
In many structured systems, such as~\eqref{eq:stationary_eqs},
this upper bound is much larger than the actual number of solutions.  
A refinement of this is the multi-homogeneous bound, which will be used
below to obtain a sharp upper bound of $2N$ for~\eqref{eq:stationary_eqs}.

From such a bound, one constructs another system that has exactly that 
many isolated nonsingular solutions which is easy to solve.
A homotopy from this system to the given system is constructed
which defines solution paths.  The endpoints of convergent paths
form a superset of the isolated solutions of the given system.

\subsection{Upper bound on the number of stationary points}
The B\'ezout bound for the stationary equations~\eqref{eq:stationary_eqs} of the $2$-spin model is $2^{N+1}$.  However, due to the
structure of the system which has a natural partition of the variables,
namely $\textbf{x}$ and $\lambda$, this B\'ezout count is far from sharp.
In fact, a well-known bound on the maximum number of \textit{real} stationary points 
is $2N$ \cite{Fyodorov-topo-triv2013}, which can be obtained, for example,
by taking $\textbf{h} = 0$.  The following shows that $2N$ is also
a sharp upper bound on the number of \textit{complex} stationary points
derived via a $2$-homogeneous B\'ezout bound.

The $2$-homogeneous bound arises from the natural partition of the variables,
with the first group consisting of the $N$ variables arising from $\textbf{x}$ 
and the second group being $\lambda$.  To compute this bound, we first need to 
find the degrees of the polynomials which respect to each group, in
this case, called the \textit{bidegree} of each polynomial.
The first $N$ polynomials in~\eqref{eq:stationary_eqs} have
bidegree $(1,1)$ since they are linear in $\textbf{x}$ and linear in $\lambda$.  
The last polynomial has bidegree $(2,0)$ since it is quadratic in~$\textbf{x}$
and $\lambda$ does not appear.  

Computing the $2$-homogeneous bound now turns into a combinatorial problem.
In particular, one needs to determine all the ways in selecting $N$ nonzero entries 
in the first spot and $1$~nonzero entry in the second spot.  
Here, $N$ and $1$ correspond to the dimensions of the spaces,
i.e., $\textbf{x}\in\CC^N$ and $\lambda\in\CC$, respectively.  
The bound is simply the sum over the products of the corresponding entries.
In particular, since the last polynomial has bidegree $(2,0)$
and the other $N$ polynomials have bidegree $(1,1$),
the $2$-homogeneous bound is simply $2$ times the number of ways
of selecting $N-1$ items out of a total of $N$ items, i.e., $2N$.  

Since there is a system which has $2N$ real solutions, i.e., taking $\textbf{h} = 0$,
it follows that, with probability $1$, 
\eqref{eq:stationary_eqs} has exactly $2N$ complex solutions.
Therefore, the $2$-homogeneous bound is (generically) sharp.
That is, from a corresponding start system with precisely $2N$ solutions,
there is a bijection, defined by the solution paths of the homotopy, 
between the $2N$ solutions of the start system
and the $2N$ solutions of each system that corresponds to the 
selected random data.  

We obtained the data via parallel computing which is based on the 
independence of solving each random instance 
and the independence of tracking each of the $2N$ paths.  
In particular, we solved using {\tt Bertini} \cite{Bertini,BHSW13}
on a cluster of $9$ processors, 
each with $8$ cores running at $2.3$ GHz.  

\section{Alternative reformulation via non-Hermitian matrices}
\label{sec:nonherm_method}
Although the NPHC method described in the previous section applies quite generally
to solving systems of multivariate polynomial equations,
we can exploit the structure of the $2$-spin spherical model
to develop another solving approach.
This method is based on non-Hermitian random matrices, which are matrices $A$ such that $A^{\mathrm{T}} \neq A$, that was suggested in~\cite{Fyodorov-topo-triv2013} but has not yet been exploited for numerical purposes. 

The first step is to note that after diagonalizing the GOE matrix $H$, the stationarity condition \eqref{eq:stationary_eqs} can be solved:
\begin{equation}
\label{eq:solution}
\textbf x^{*} = \sum_{j=1}^{N}\tilde{\textbf x}_{j}\textbf u_{j}, \qquad \tilde{\textbf x}_{j} = \frac{\tilde{\textbf h}_{j}}{\tilde{\lambda}-\lambda_{j}}
\end{equation}
where $\tilde{\textbf h}_{j} =\textbf h^{\mathrm{T}}\textbf u_{j}$ and $\textbf u_{j}$ are the sequence of orthonormal eigenvectors of $H$ with corresponding eigenvalues $\lambda_{1} < \lambda_{2} < \ldots < \lambda_{N}$ and $\tilde{\lambda} = -2\lambda$.

Next, we have to obtain an equation for $\tilde{\lambda}$. From the spherical constraint $\|\textbf x^{*}\|^{2}=N$, formula \eqref{eq:solution} gives the condition $\textbf h^{\mathrm{T}}(H-\tilde{\lambda})^{-2}\textbf h=N$. This is equivalent to the determinantal equation $\mathrm{det}((H-\tilde{\lambda})^{2}-N^{-1}\textbf h\textbf h^{\mathrm{T}})=0$. Finally, using the well-known formula for the determinant of a block matrix, we see that $\tilde{\lambda}$ satisfies \eqref{eq:stationary_eqs} if and only if $\tilde{\lambda}$ is a \textit{real} eigenvalue of the following non-Hermitian block matrix
\begin{equation}
A = \begin{pmatrix} H & N^{-1}\textbf h\textbf h^{\mathrm{T}}\\
I_{N} & H \end{pmatrix} \label{eq:nonhermitianA}
\end{equation}
where $I_{N}$ is the $N \times N$ identity matrix. Notice that when $\textbf{h}=0$, $A$ has the same eigenvalues of $H$, and there are $2N$ stationary points. 
Then, the external field $\textbf{h} \neq 0$ breaks the symmetry of $A$ and pushes a non-trivial fraction of the eigenvalues into the complex plane.

In summary, we see that to compute the real solutions of \eqref{eq:stationary_eqs}, 
it is sufficient just to calculate the real eigenvalues of the matrix $A$ to obtain 
all possible values of $\tilde{\lambda}$. The total number of such real eigenvalues gives the total number of stationary points. 
Then, the positions of the stationary points can be obtained by inserting all possible real values of $\tilde{\lambda}$ into \eqref{eq:solution} 
to obtain $\textbf x^{*}$. The energy of each stationary point can then be computed from \eqref{2hamilton}. 
The numerical results of such a procedure %sp - is -> are 17/09
are described in Section~\ref{sec:nonherm_numerics}.
We also compare with the general purpose NPHC method from the previous section.

To calculate the mean and the variance, as well as the frequency distribution of the total number of stationary points, it suffices to generate enough realizations of the matrix $A$ in \eqref{eq:nonhermitianA} and to count the real eigenvalues for each realization. This was done by setting up the block matrix $A$ in Matlab and each time computing the eigenvalues using the built-in function {\tt eig}. 
The number of realizations used for the data presented here was 100,000
except for $N=200$ in which only 50,000 realizations were used.
%Since each $\lambda$ gives rise to a stationary point of the energy \ref{2hamilton}, we see that the number of real eigenvalues of $A$ is identical to the number of stationary points. Further, their locations and the corresponding energies may then be directly evaluated using formula \ref{eq:solution}.
\section{Results}
\label{s:results}
\label{sec:nonherm_numerics}
In the following we present the results of the computations based on 
the numerical approaches outlined above.
When investigating the behavior of the real solutions, the
non-Hermitian matrix method is preferred due to the speed of the computation.  
We did, however, verify the results matched computations using NPHC method.
When investigating the behavior of both the 
real and imaginary parts of the Hamiltonian, 
this involved using the NPHC method.  

%In Figure \ref{fig:no_of_SPs-gamma} we plot the mean number of stationary points against increasing values of $\gamma$ for various $N$. The plot shows a rapid convergence to the limiting $N = \infty$ curve given by \ref{eq:mean.real.sols.asymp.gamma}. In the $\kappa$-regime, the convergence is a little more delicate, especially near $\kappa=0$, but still shows convergence to the known theoretical results. 

%One advantage of the present method is that we were able to generate enough realizations to get an accurate picture of the variance, especially in the $\gamma$ regime. The latter quantity is plotted in figure \ref{fig:var_no_of_SPs-gamma} and also appears to show convergence to a limiting curve after normalizing the total number of critical points by $N^{-1/2}$. In this case the limiting curve has not been derived before by any theoretical methods. We also show the analogous plots for the $\kappa$ regime in Figure \ref{fig:var_no_of_SPs-kappa} which seems to require the same normalization.

%In Figure \ref{fig:gamma-histogram} we plot the probability histograms for the distribution of the number of stationary points for $N=100$ and various values of $\gamma$ with $100,000$ realizations. The plot shows the same qualitative features as the $N=75$ figure \ref{fig:frequencies_of_real_solutions} computed from the NPHC method.

\subsection{Mean number of stationary points}
\begin{figure}
\includegraphics[scale=0.4]{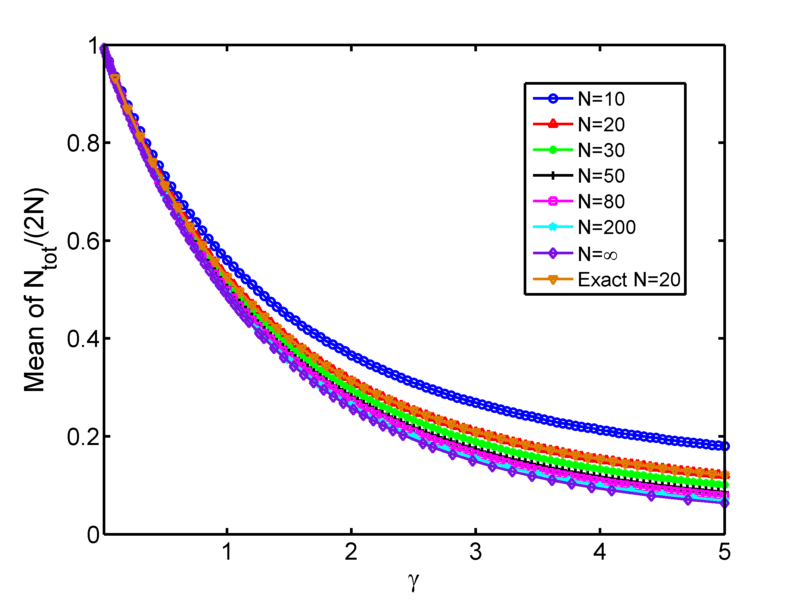} 
\caption{Mean number of stationary points as a function of $\gamma$.}
\label{fig:no_of_SPs-gamma}
\end{figure}

\begin{figure}
\includegraphics[scale=0.4]{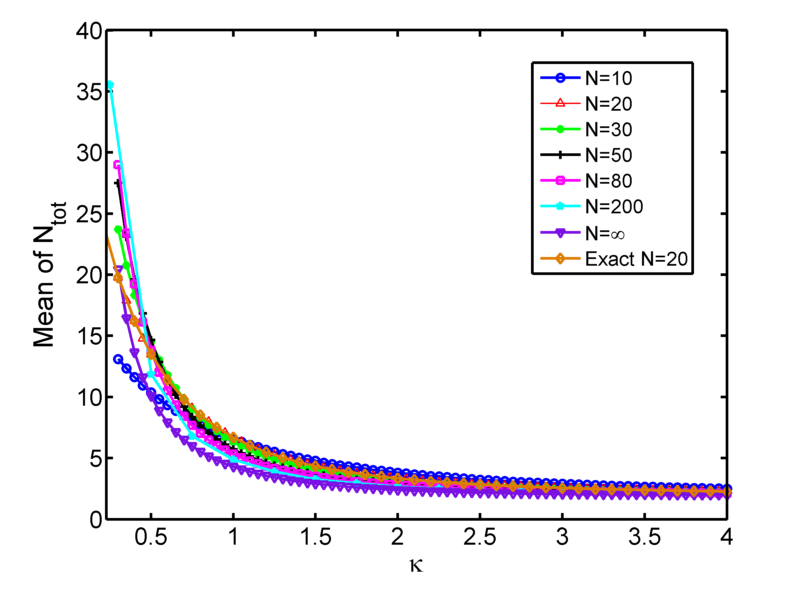}
\caption{Mean number of stationary points as a function of $\kappa$.} 
\label{fig:no_of_SPs-kappa}
\end{figure}

In Figures \ref{fig:no_of_SPs-gamma} and \ref{fig:no_of_SPs-kappa}, the average number of real solutions are shown as a function of $\gamma$ and $\kappa$, respectively. Each point in the plots represents the average over 100,000 samples. Numerical results from the non-Hermitian eigenvalue problem \eqref{eq:nonhermitianA} are plotted for several different values of the dimension $N$, together with the theoretical results in the asymptotic limit from \eqref{eq:mean.real.sols.asymp.gamma} and \eqref{eq:mean.real.sols.asymp.kappa} and also with the exact expression from \eqref{eq:mean.real.sols.exact}. 

%\begin{figure}
%\subfigure[]{\includegraphics[scale=0.4]{gamma-mean.eps} \label{fig:no_of_SPs-gamma}}
%\subfigure[]{\includegraphics[scale=0.4]{kappa-mean.eps} \label{fig:no_of_SPs-kappa}}
%\caption{}
%\end{figure}

In Fig. \ref{fig:no_of_SPs-gamma}, the finite $N$ numerical results show a qualitatively similar trend to the asymptotic results, approaching this in a relatively fast rate as $N$ grows. The results are also compared with the exact analytic formula (\ref{eq:mean.real.sols.exact}) for a fixed size $N=20$. The numerical calculations agree excellently with the analytical expressions. The same observations are valid for Fig. \ref{fig:no_of_SPs-kappa} which shows the results for the $\kappa$ regime. Here ${\mathcal N}(\kappa) \to 2$ for large $\kappa$, which is the limiting regime of topology trivialization as described above. In summary, these results show both
the correctness of the analytical approaches for computing the mean number of stationary points in the GOE ensemble, and also the correctness of the numerical calculations from the non-Hermitian eigenvalue problem \eqref{eq:nonhermitianA}.

\subsection{Variance of the number of real stationary points}
While it is often possible to compute analytical expressions for the mean number of real solutions of a random system of equations, 
obtaining analytical expressions for the variances or higher order moments of the distribution is often a very difficult task, if not impossible. Indeed, for the $2$-spin spherical model, analytical expressions for the variance for both finite $N$ and $N \to \infty$ are completely unknown. It is here where numerical methods can be most useful.

By means of the non-Hermitian matrix \eqref{eq:nonhermitianA}, we can find all the real solutions for each sample of the $2$-spin model, and then we can straightforwardly compute the variance of the number of real solutions. This quantity, which is a measure of the fluctuations of the mean number of real solutions, is of particular relevance as it gives information on the occurrence of real versus complex solutions of the system of equations in the different regimes.

\begin{figure}[htp!]
\includegraphics[scale=0.4]{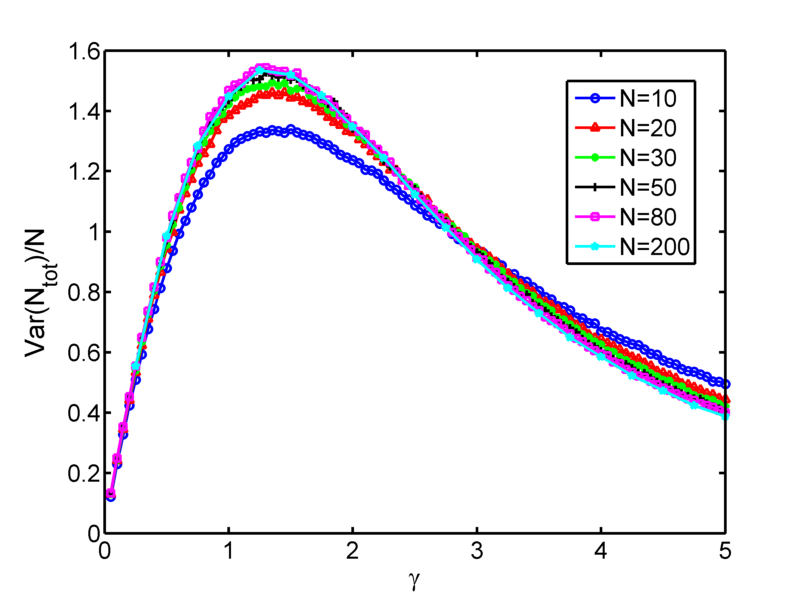}
\caption{Variance of the number of stationary points as a function~of~$\gamma$.}
\label{fig:var_of_SPs-gamma}
\end{figure}

\begin{figure}
\includegraphics[scale=0.4]{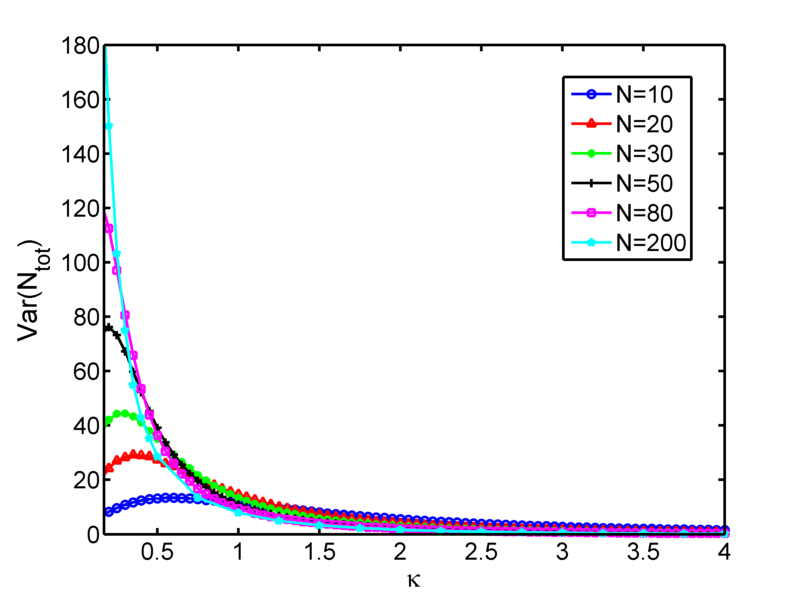}
\caption{Variance of the number of stationary points as a function~of~$\kappa$.}
\label{fig:var_of_SPs-kappa}
\end{figure}

The variance as a function of $\gamma$ and $\kappa$ for different values of $N$ is plotted in Figures \ref{fig:var_of_SPs-gamma} and \ref{fig:var_of_SPs-kappa}, respectively. In Figure \ref{fig:var_of_SPs-gamma}, as we increase through higher values of $N$, the variance shows a clear convergence to a well defined limiting curve, confirming our normalization of $N_{tot}$ by $N^{-1/2}$ in this context. An important open problem is to provide a theoretical justification for this normalization and the resulting limiting curve.  In the $\kappa$ regime,
shown in Figure~\ref{fig:var_of_SPs-kappa}, 
the number of stationary points is characterized by large fluctuations near the origin $\kappa=0^{+}$ which are quickly suppressed for increasing values of~$\kappa$.

%\begin{figure}[ht!]
%\includegraphics[scale=0.35,angle=270]{gamma-stddev.eps}\\
%\includegraphics[scale=0.35,angle=270]{kappa-stddev.eps}
%\caption{}
%\label{fig:variances}
%\end{figure}

\subsection{Frequencies of the no. of stationary points}
\begin{figure}[ht!]
\includegraphics[scale=0.4]{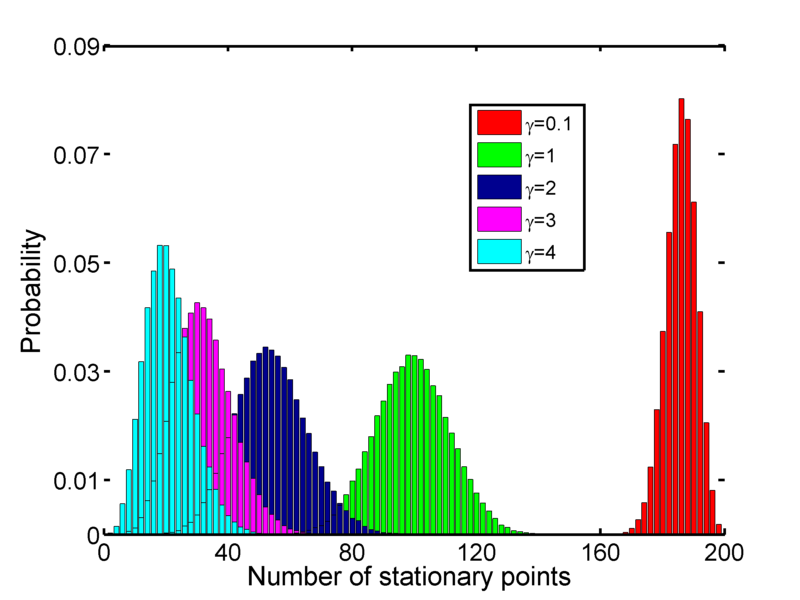}
\caption{Probability densities of the number of stationary points for different values of $\gamma$.}
\label{fig:gamma-histogram}
\end{figure}

Going beyond the mean and variance, we can also obtain the full distribution of the number of stationary points. The results are plotted in Figure \ref{fig:gamma-histogram} in the $\gamma$ regime for $N=75$. The plots were generated from 100,000 realizations of the matrix $A$ in equation $\eqref{eq:nonhermitianA}$. For increasing values of $\gamma$, we note the spread of the distribution behaving in accordance with the variance plot in Figure \ref{fig:var_of_SPs-gamma}. 
As with the variance, there is not yet any analytic results about the full distribution of the number of stationary points. Its theoretical investigation may be of broader interest to practitioners of random matrix theory, as the number of real eigenvalues were investigated by several authors when the underlying matrix is composed of independent, identically distributed entries \cite{TaoVu13} or satisfies invariance \cite{EKS93} with respect to the action of an appropriate compact group. In these simpler cases, it was proven that the fluctuations of the real eigenvalue count are Gaussian when $N \to \infty$, with mean and variance of order $\sqrt{N}$. In contrast, our study shows that for the matrix $A$ in the $\gamma$-regime, the real eigenvalues instead have mean and variance of order $N$. %The random variables plotted in Figure \ref{fig:frequencies_of_real_solutions} can be realized as a linear statistic on the real eigenvalues of the matrix $A$:
%\begin{equation}
%\label{ls}
%LS := \sum_{j=1}^{2N}\chi_{\mathbb{R}}(\lambda_{j})
%\end{equation}
%where $\chi_{\mathbb{R}}$ denotes the indicator function on the real line. For many classes of random matrices, it was shown that after centering and normalizing, statistics of the form \eqref{ls} converge in distribution to a Gaussian variable. 

%\begin{figure}[ht!]
%\includegraphics[scale=0.35,angle=270]{gamma-histogram-N75.eps}
%  \caption{}
%  \label{fig:frequencies_of_real_solutions}
%\end{figure}

\subsection{Distribution of global minima}

%In Figure \ref{fig:dist-of-minima-gama2-N50} we show the (normalized) distribution of ground state energies for
%$N=50$ and $\gamma=2$, together with the analytical prediction for small deviations from the typical value of the
%energy, which is a gaussian distribution \cite{Fyodorov-topo-triv2013}:
%\beq
%P(E)dE \approx \frac{1}{\sqrt{N\Gamma \pi}}\exp{\left[-\frac{(E-E_{typ})^2}{N\Gamma}\right]},
%\eeq
%with $\Gamma=\sigma^2$.
%This approximation should be valid for $N^{1/2} \ll |E-E_{typ}| \ll N$. For other values in the $\gamma$ regime the
%results are similar. Note that because of the definition of $\Gamma$, in the $\gamma$ regime the Gaussian is
%independent of $N$.

\begin{figure}[ht!]
\includegraphics[scale=0.35]{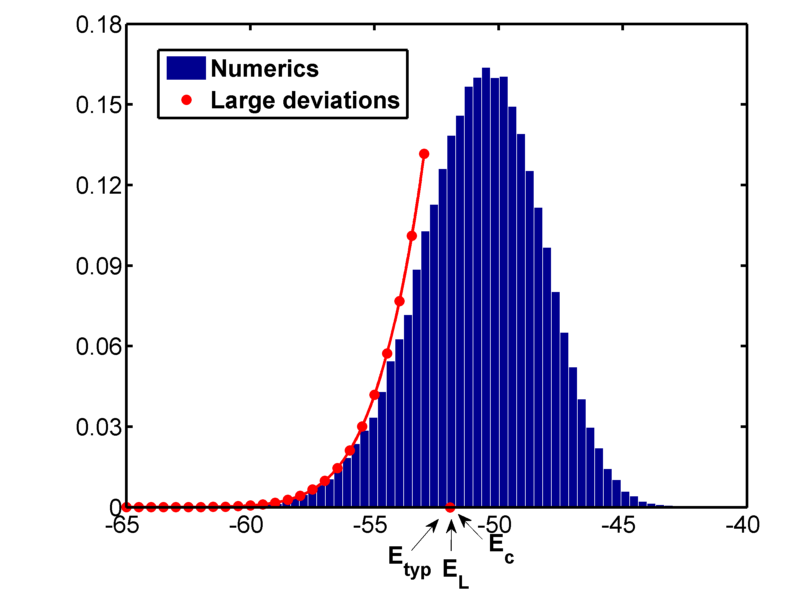}
  \caption{Probability density of $E_{\min}$ for $\gamma=2$ and $N=50$.}
  \label{fig:dist-of-minima-gama2-N50}
\end{figure}

In order to obtain the distribution of the global energy minimum with our methods, one simply takes the obtained values of the Lagrange multipliers (namely, the eigenvalues of the matrix $A$ in \eqref{eq:nonhermitianA}) and inserts the results into \eqref{eq:solution}. Then, numerically, it's a simple task to evaluate the energy $E_{h}(\mathbf{x})$ at the $2N$ critical points and minimize over all outputs. The corresponding probability histogram is depicted in Figure \ref{fig:dist-of-minima-gama2-N50} for $N=50$, $J=1$, $\gamma=2$ with 100,000 realizations.

\begin{figure}[ht!]
\includegraphics[scale=0.35]{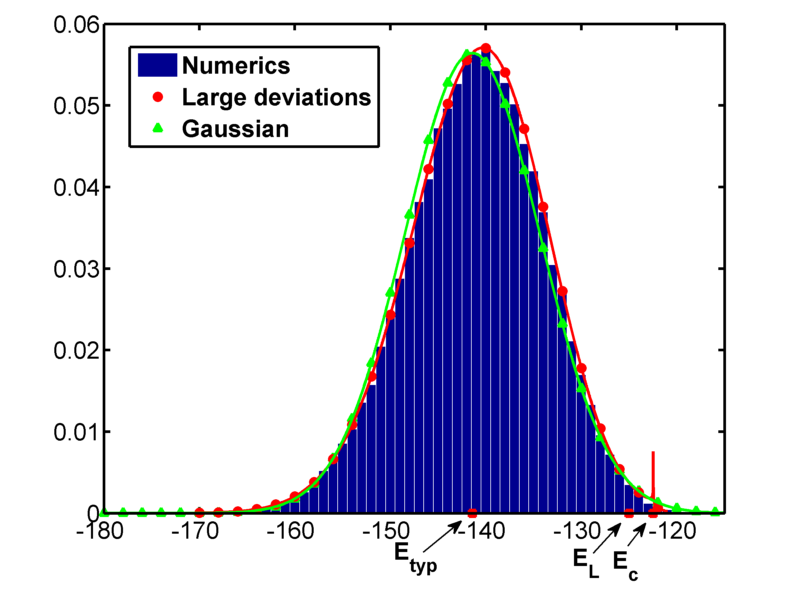}
  \caption{Probability density of $E_{\min}$ for $\sigma=1$ and $N=100$.}
  \label{fig:dist-of-minima-sig1-N100}
\end{figure}

\begin{figure}[ht!]
\includegraphics[scale=0.35]{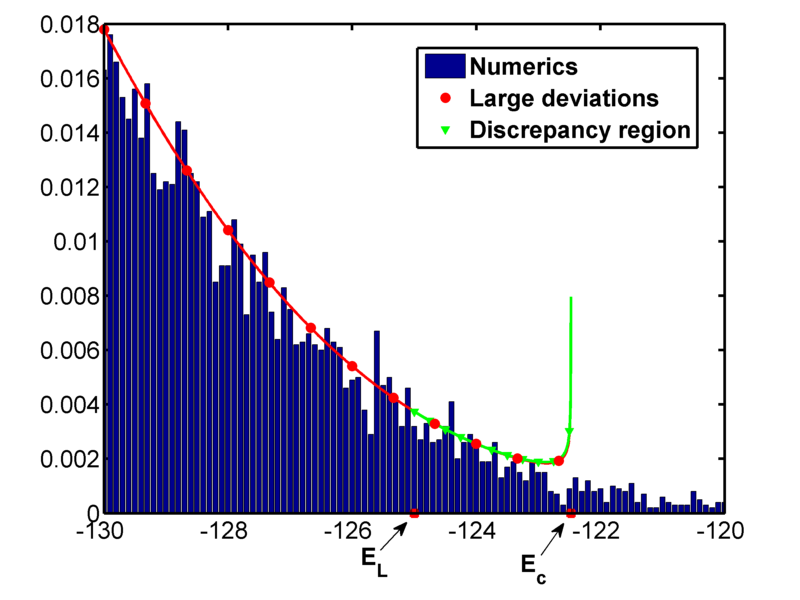}
  \caption{Probability density of $E_{\min}$ near the critical energy, again with $\sigma=1$ and $N=100$.}
  \label{fig:dist-of-minima-sig1-N100-comp}
\end{figure}

\begin{figure}[ht!]
\includegraphics[scale=0.35]{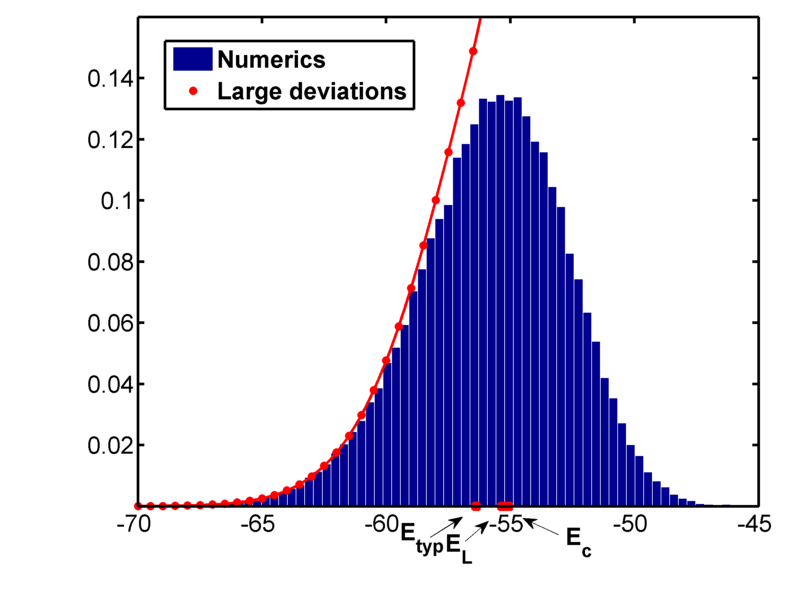}
  \caption{Probability density of $E_{\min}$ for $\kappa=1$ and $N=50$.}
  \label{fig:dist-of-minima-kap2-N50}
\end{figure}

\begin{figure*}[ht!]
\centering
\begin{minipage}[t]{.4\textwidth}
\includegraphics[scale=0.35]{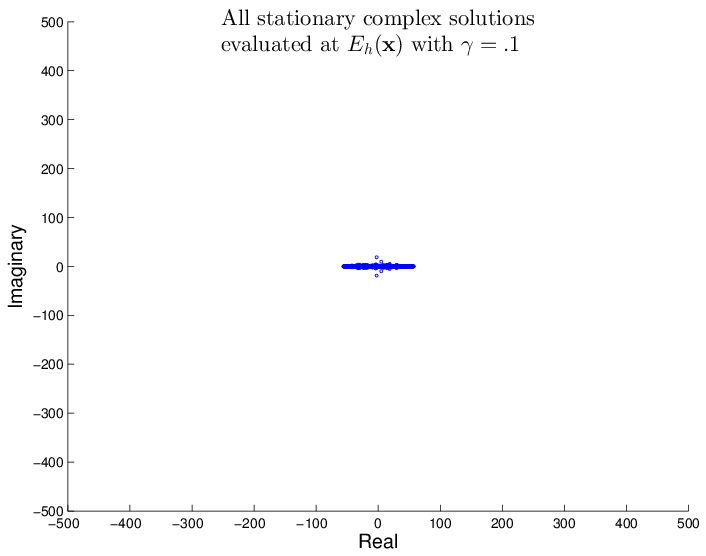}
%\caption{}
\label{fig:pot1}
\end{minipage}\qquad
\begin{minipage}[t]{.4\textwidth}
\includegraphics[scale=0.35]{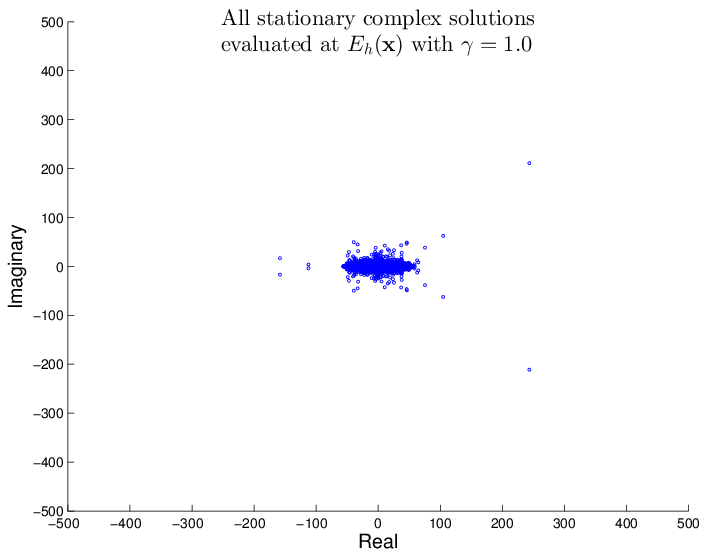}
%\caption{}
\end{minipage}
%\begin{minipage}[t]{.4\textwidth}
%\includegraphics[scale=0.35]{png/pot_eval_gam_20.png}
%\caption{}
%\end{minipage}\qquad
\begin{minipage}[t]{.4\textwidth}
\includegraphics[scale=0.35]{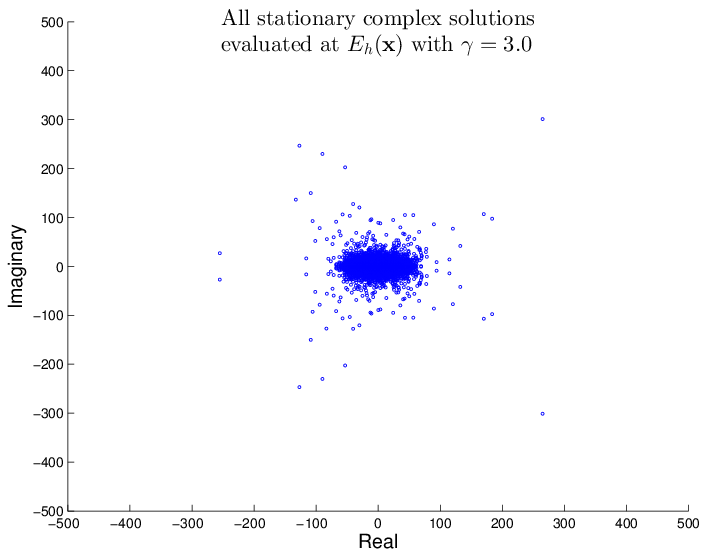}
%\caption{}
\end{minipage}
%\begin{minipage}[t]{.4\textwidth}
%\includegraphics[scale=0.35]{png/pot_eval_gam_40.png}
%\caption{}
%\end{minipage}\qquad
\begin{minipage}[t]{.4\textwidth}
\includegraphics[scale=0.35]{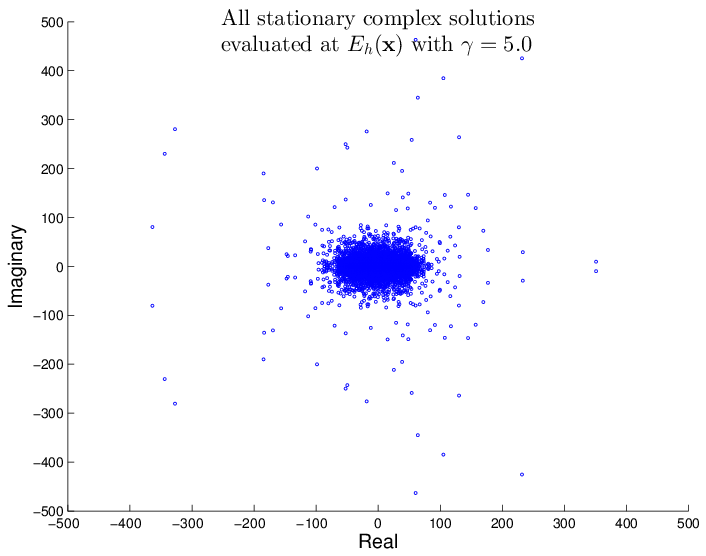}
%\caption{}
\end{minipage}
\begin{minipage}[t]{.4\textwidth}
\includegraphics[scale=0.35]{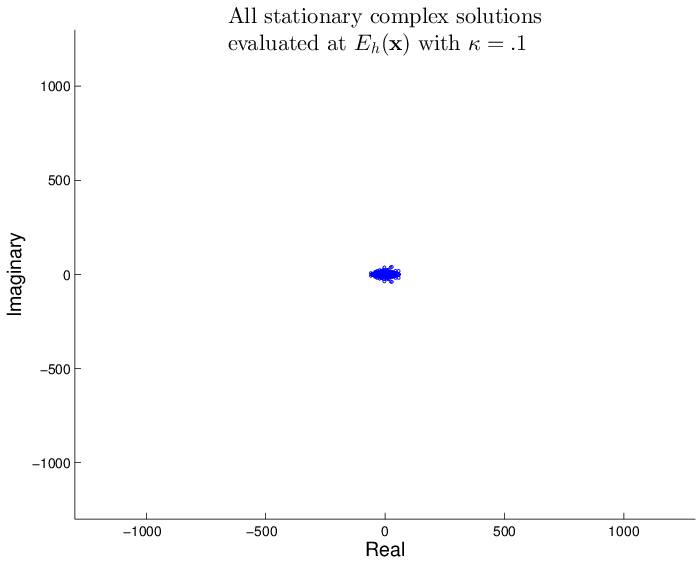}
%\caption{}
\end{minipage}\qquad
\begin{minipage}[t]{.4\textwidth}
\includegraphics[scale=0.35]{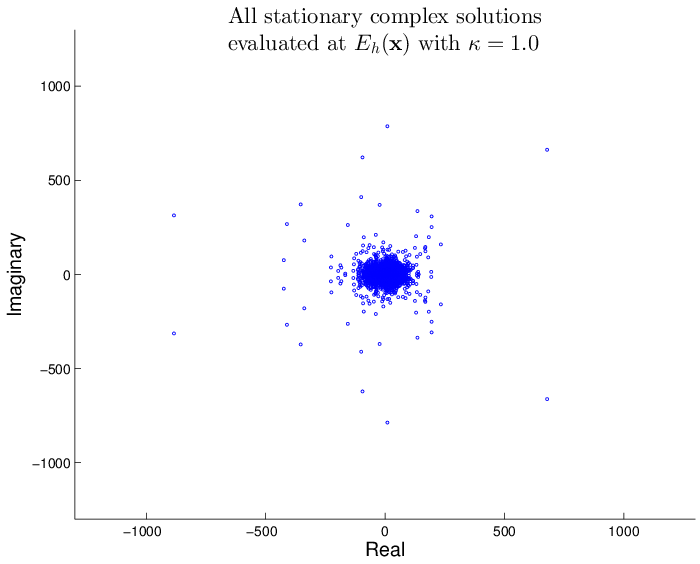}
%\caption{}
\end{minipage}
\begin{minipage}[t]{.4\textwidth}
\includegraphics[scale=0.35]{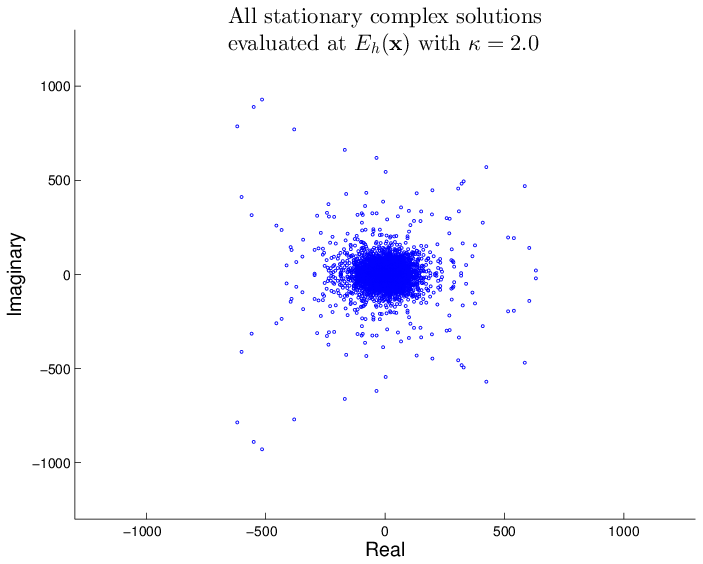}
%\caption{}
\end{minipage}
%\begin{minipage}[t]{.4\textwidth}
%\includegraphics[scale=0.35]{png/pot_eval_kap_30}
%\caption{}
%\end{minipage}\qquad
\begin{minipage}[t]{.4\textwidth}
\includegraphics[scale=0.35]{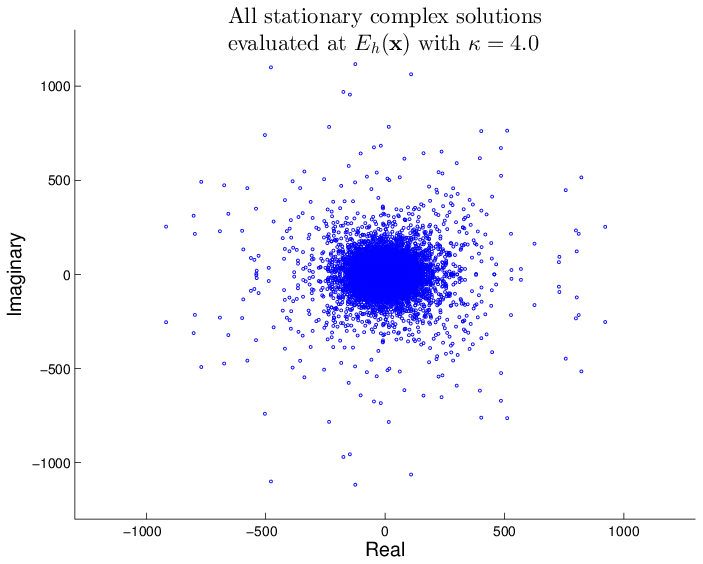}
%\caption{}
\end{minipage}
\caption{Plots of the real and imaginary parts of the energy function $E_{h}(\mathbf{x})$ evaluated at the complex stationary points for different values of $\gamma$ and $\kappa$.}
\label{fig:pots}
\end{figure*}

The statistical properties of the ground state energy of the 2-spin spherical model were investigated analytically in \cite{Fyodorov-topo-triv2013} and later in \cite{DemZei14}. In \cite{Fyodorov-topo-triv2013}, a large deviations asymptotic expression for the probability density function of $E_{\min}$ was derived, valid up to a critical value of the energy $E_{c} := -N\sqrt{\frac{1+2\sigma^{2}}{1+\sigma^{2}}}$ and depending on the parameter $E_{\mathrm{typ}} = -N\sqrt{1+\sigma^{2}}$, the typical value of $E_{\min}$. Recently the corresponding rate function was obtained rigorously in \cite{DemZei14}, revealing a surprising difference with the one obtained in \cite{Fyodorov-topo-triv2013}. Specifically, it was shown in \cite{DemZei14} that there is a different critical parameter $E_{{L}} := -N\left(1+\frac{\sigma^{2}}{2(1+\sigma^{2})}\right)$ for which the two rate functions \textit{disagree} on the interval $[E_{{L}},E_{c}]$.

In Figure \ref{fig:dist-of-minima-gama2-N50}, we plot the large deviations functional in \cite{Fyodorov-topo-triv2013} that was also proved rigorously in \cite{DemZei14}. The results show a good consistency between the two approaches in the regime of validity of large deviations $E \ll E_{{c}}$. The values of $E_{{c}}$, $E_{{L}}$ and $E_{{typ}}$ are almost identical here.

On the other hand, if we consider the regime of topology trivialization, where $\sigma>0$ is fixed, we get an almost perfect agreement with large deviations, see Figure \ref{fig:dist-of-minima-sig1-N100}, where we set $\sigma = 1$, $J=1$ and $N=100$. The reason seems to be that for fixed $\sigma$, the threshold $E_{{c}}$ moves far out into the right tail of the distribution, giving a wider range of validity. The triangles show the Gaussian
\begin{equation}
P(E) \propto \mathrm{exp}\left(\frac{(E-E_{\mathrm{typ}})^{2}}{\sigma^{2}N}\right)
\end{equation}
giving a good approximation to the tails of the distribution \cite{Fyodorov-topo-triv2013}. 

For $\sigma = 1$, the critical parameters also begin to separate out more and one can ask how the two large deviations expressions differ on $[E_{{L}},E_{{c}}]$. As seen in Figure \ref{fig:dist-of-minima-sig1-N100-comp}, this difference is very small and is hard to detect numerically. Below $E_{{c}}$, the triangular data points are based on the rigorous large deviations formula in \cite{DemZei14} and circles the one in \cite{Fyodorov-topo-triv2013}. At the level of rate functions, their difference is upper bounded by $10^{-4}$ on the interval $[E_{{L}},E_{{c}}]$. Away from this interval, the two expressions are identical \cite{DemZei14}. The plot also shows that as one approaches $E_{{c}}$ the pre-exponential factor in \cite{Fyodorov-topo-triv2013} diverges and should be replaced by a different expression beyond the threshold $E_{{c}}$.

Finally, we plot the results for the $\kappa$-regime in Figure~\ref{fig:dist-of-minima-kap2-N50}. Now, the large deviation expressions gives an agreement somewhere in between the last two regimes, as expected from the fact that $\sigma_{\gamma} \ll \sigma_{\kappa} \ll 1$, where $\sigma_{\gamma}$ and $\sigma_{\kappa}$ denote the $\sigma$ values corresponding to the $\gamma$ and $\kappa$ regimes respectively.

%\begin{figure}[ht!]
%\includegraphics[scale=0.35,angle=270]{min-dist-N50-gama2.eps}
%  \caption{}
%  \label{fig:dist-of-minima-gama2-N50}
%\end{figure}

%In Figure \ref{fig:dist-of-minima-kappa1-N50} we show the probability density of ground state energies for $N=50$ and $\kappa=1$. In this case, due to the different
%dependence of $\Gamma$ with $N$, the gaussian approximation becomes pourer as $\kappa$ grows.

%\begin{figure}[ht!]
%\includegraphics[scale=0.35,angle=270]{min-dist-N50-kappa1.eps}
%  \caption{}
%  \label{fig:dist-of-minima-kappa1-N50}
%\end{figure}

\subsection{Complex Stationary Points}
As stated before, the NPHC method finds all complex solutions of
\eqref{eq:stationary_eqs}.
Since, with probability $1$, there are always $2N$ complex solutions for any random sample, only the number of real solutions varies with $\gamma$ and $\kappa$. In other words, while increasing $\gamma$
and $\kappa$, some of the real stationary points become complex solutions.
One way of studying this phenomenon is by plotting real vs imaginary parts of $E_{h}(\textbf{x})$, see Figure \ref{fig:pots}.
The plots show that at small $\gamma$ and~$\kappa$, the imaginary part of $E_{h}(\textbf{x})$ evaluated at all the $2N$ complex stationary 
points is zero. As the parameters increase, the imaginary parts of $E_{h}(\textbf{x})$ increases meaning that some of the real solutions became nonreal.

%\begin{figure}[ht!]
%\centering
%
%\begin{subfigure}
%\includegraphics[scale=0.35]{pot_eval_gam_1.eps}
%  \caption{}
%\label{fig:Re_vs_Im_V_gamma_1}
%\end{subfigure}
%
%\begin{subfigure}
%\includegraphics[scale=0.35]{pot_eval_gam_10.eps}
%  \caption{}
%\label{fig:Re_vs_Im_V_gamma_10}
%\end{subfigure}
%
%\end{figure}

\section{Discussion and Conclusion}
Exploring potential energy landscapes of various models arising in physics and chemistry is a very active area of research in different fields
of science and mathematics. Recently, a curious feature of the %sp - potentials -> potential 17/09
potential energy landscapes of a class of statistical mechanics models
has been observed, namely, topology trivialization: while varying one or more parameters of the potential, either continuously or varying
the variance of the random distribution the parameter values are drawn from, the mean number of real stationary points of the potential varies from
$O(1)$ to $O(N)$ or even higher. In the former case, the topology of the $N$-dimensional landscape can be viewed as being \textit{trivialized}.
These two phases are shown to be related to phase transitions of the systems. 
In this work we have done a numerical study of the topology trivialization scenario in the $2$-spin spherical model. 
While the mean number of real stationary points 
can be computed
analytically using random matrix theory tools, computing other quantities such as the variance of the number of real stationary points and the full distribution are prohibitively difficult for current \mbox{analytical~computation~techniques.} 
%sp - removed "not to say" above - it sounded a bit clunky. Also replaced "later" with "latter" in the previous paragraph. - NS 19/09
%Here, we have concentrated on the $2$-spin model. The parameters to be varied here are $\gamma$ and $\kappa$ which are functions of the 
%variances of the Gaussian distributions from which the disorders and the external magnetic field are drawn from.

We used two numerical methods, namely, the numerical polynomial homotopy continuation (NPHC) method and non-Hermitian matrix method.
One first translates the problem of finding stationary points
into an algebraic geometry problem of solving a system of polynomial equations.
This interpretation yields an upper bound on the number of complex solutions, 
namely $2N$ which is equal to the known upper bound on the 
number of \textit{real} solutions for this system. 
In fact, $2N$ is equal to the number of complex solutions, with probability~$1$,
and only the number of real solutions varies with each instance.
Hence, we have found a more general result for the number of 
solutions of the $2$-spin model. 

The second method, though apparently only specific to the $2$-spin case, works more efficiently in this case by finding all the real solutions for a given random
instance and hence giving an opportunity to reach much higher dimension $N$ and sample size. The method does not find complex solutions which were analyzed
using the NPHC method.

With the two powerful methods at our disposal, we first reproduced the analytical predictions on the mean number of 
real solutions with an excellent agreement. We also addressed the issue of fluctuations of the number of solutions, 
showing that for the $\gamma$-regime, the variance of the number of critical points is of order $N$ as $N \to \infty$. 
To show this analytically seems to us an important open problem. 
Little is known in general about fluctuations of the number of critical points in random Gaussian fields, although in a different context results in this direction were obtained in \cite{KleAga12}.

We also investigated statistics of the global energy minimum $E_{\min}$. When $\sigma>0$ is fixed and large enough that 
$E_{c} \gg E_{\mathrm{typ}}$ (corresponding to the regime of topology trivialization), our findings give a strong agreement with the heuristic arguments in \cite{Fyodorov-topo-triv2013}. Remarkably, it seems that in this regime, the entire distribution of $E_{\min}$ yields precise agreement with the large deviations expression in \cite{Fyodorov-topo-triv2013}. In the $\gamma$ and $\kappa$ regimes, the agreement with large deviation theory is limited to the left tail of the distribution. The reason seems to be that when $\sigma \to 0$, the critical energy threshold $E_{c}$ moves further into the bulk of the distribution and we know that the pre-exponential factors from \cite{Fyodorov-topo-triv2013} are not valid if $E > E_{c}$. Analytical understanding of the statistics of $E_{\min}$ in the right tail for the $\gamma$ and $\kappa$ regimes therefore remains an outstanding issue. 

We note that the topology trivialization phenomenon, at least in the simple case of continuously varying parameters, 
shares a deep connection with Catastrophe theory, which is now absorbed in a more general mathematical framework of singularity theory and bifurcation
theory.
From Catastrophe theory, it is known that varying the parameters of the potential continuously the real stationary points may appear or disappear, 
or change
their stability properties \cite{wales2001microscopic,bogdan2004new}. In Refs.~\cite{Kastner:2011zz, Mehta:2012qr,Mehta:2014comm}, it was observed 
that while continuously varying the parameter of the 
two-dimensional nearest neighbor $\phi^4$ model, some of the real stationary points would merge to become complex solutions and vice versa.

The fact that the topology trivialization occurs when varying the variance of the random distributions from which the parameters are drawn,
rather than varying the parameters themselves, makes such a description more subtle. In the present work, however, we have observed
that a similar phenomenon of real stationary points transforming to complex and vice versa is occurring in the $2$-spin model too when varying
$\gamma$ and $\kappa$. 

Another description of the topology trivialization phenomenon may come from our algebraic geometry interpretation of the $2$-spin model: for a 
simple system $a x^2 + b x + c = 0$, where $a,b$ and $c$ are real parameters, the discriminant $b^2-4 a c$ decomposes the 3D parameter space in to
three phases, i.e., no real roots, two distinct real roots and double roots. Thus, the number of real solutions goes from the highest possible to 
zero. Similarly, a discriminant can be defined for multivariate polynomials case and a similar classification of the parameter space
based on the number of real solutions can be worked out using the so-called discriminant variety method 
\cite{gel1994fand,lazard2007solving,hanan2010stability}. From this, one can study the topology 
trivialization fairly straightforwardly for the case of continuously varying parameters. However, the case of varying variances of the random
distributions of the parameters is still subtle and largely unexplored even from the Mathematics point of view.

Thus, we anticipate that our results will merge the topology trivialization phenomenon 
with the emerging mathematical areas called Statistical Topology, 
or perhaps inspire a new subbranch that may be called
\textit{statistical catastrophe theory} or \textit{stastical discriminant variety}. 

We also note that for higher $N$, numerical instabilities become profound when finding stationary points of the $p$-spin model using the above 
numerical
methods. To resolve this issue, one can employ, for example, Smale's alpha theorem to certify if a numerical approximate is provably within the
quadratic convergence region of the nearby exact root. Combining this certification with the NPHC method then gives a result equivalent to the exact
result for each random instance \cite{Mehta:2013zia,Mehta:2014gia}. In the future, we plan to use this combination to prove concrete results for 
higher values of $N$.

\acknowledgments
J.D.H.\ and D.M.\ were supported by a DARPA Young Faculty Award.  
J.D.H.\ and M.N.\ were partially supported by 
the Simons Institute for the Theory of Computing
with NIMS at CAMP also providing support to M.N.
D.A.S.\ acknowledges partial support from CNPq, Brazil. 
We thank Yan Fyodorov and Michael Kastner for their 
critical remarks and
feedback at various stages of this work.

\end{document}